\documentclass[preprint]{aastex}
\input epsf
\newcommand{\reff}{\mbox{$r_{\rm eff}$}}
\newcommand{\rmax}{\mbox{$r_{\rm max}$}}

\newcommand{\msun}{\mbox{$M_{\odot}$}}
\newcommand{\mvir}{\mbox{$M_\mathrm{vir}$}}

\begin{document}
\title{Dynamical Mass Estimates for Five Young Massive Stellar Clusters 
 \footnote{Based on data obtained at the W.\ M.\ Keck Observatory, which is
   operated as a scientific partnership among the California Institute of
   Technology, the University of California, and the National Aeronautics
   and Space Administration. Also based on observations with the Hubble Space 
   Telescope, obtained at the Space Telescope Science Institute, which is 
   operated by the Association of Universities for Research in 
   Astronomy, Inc.\ under NASA contract No.\ NAS5-26555.}
}
\author{S{\o}ren S.\ Larsen
        \affil{ESO / ST-ECF, Karl-Schwarzschild-Strasse 2,
	85748 Garching b.\ Munich, Germany}
        \email{slarsen@eso.org}
\and
        Jean P.\ Brodie
        \affil{UC Observatories / Lick Observatory, University of California,
        Santa Cruz, CA 95064, USA}
	\email{brodie@ucolick.org}
\and
        Deidre A.\ Hunter
	\affil{Lowell Observatory, 1400 West Mars Hill Road, Flagstaff,
	AZ 86001, USA}
	\email{dah@lowell.edu}
}

\begin{abstract}
  We have obtained high-dispersion spectra for four massive star clusters in 
the dwarf irregular galaxies NGC~4214 and NGC~4449, using the HIRES 
spectrograph on the Keck I telescope. Combining the velocity dispersions of 
the clusters with structural parameters and photometry from images taken 
with the \emph{Hubble Space Telescope}, we estimate mass-to-light ratios and 
compare these with simple stellar population (SSP) models in order to 
constrain the stellar mass functions (MFs) of the clusters.  For all clusters 
we find mass-to-light ratios which are similar to or slightly higher than for 
a Kroupa MF, and thereby rule out any MF which is deficient in low-mass stars 
compared to a Kroupa-type MF.  The four clusters have virial masses ranging 
between $2.1\times10^5\msun$ and $1.5\times10^6\msun$, half-light radii 
between 3.0 and 5.2 pc, estimated core densities in the range 
$2\times10^3 \, \msun \, {\rm pc}^{-3}$ to 
$2\times10^5\,\msun \, {\rm pc}^{-3}$ and ages between 200 Myr and 800 Myr.
We also present new high-dispersion near-infrared spectroscopy for a luminous 
young ($\sim15$ Myr) cluster in the nearby spiral galaxy NGC~6946, which we 
have previously observed with HIRES.  The new measurements in the infrared 
agree well with previous estimates of the velocity dispersion for this 
cluster, yielding a mass of about $1.7\times10^6 \msun$. 
The properties of the clusters studied here are all consistent
with the clusters being young versions of the old globular clusters found 
around all major galaxies.  
\end{abstract}


\section{Introduction}

  The term ``super star cluster'' was probably first coined by \citet{van71}, 
who used it to describe a number of highly luminous, compact sources in the 
starburst in M~82. Similar clusters have since been found in many other
galaxies, often (but not exclusively) in starbursts and mergers. Thanks to
high resolution imaging with the \emph{Hubble Space Telescope}, it is now
clear that these clusters in many ways resemble young versions of the 
classical (old) globular clusters found around all large galaxies. This,
in turn, has fuelled expectations that these objects may provide an 
opportunity to gain first-hand insight into the formation mechanisms and
early evolution of globular clusters.

  One question which has spawned considerable debate concerns the stellar 
initial mass function (IMF) of young star clusters, and whether or not
variations exist. Such variations might have implications for cluster
survival, since clusters with a shallow IMF are more easily 
disrupted \citep{goo97} and may not be able to survive for a Hubble time.  
By definition the \emph{initial} mass function can only be observed at
the time of formation. Observations of older clusters (or stellar
populations in general) reveal a present-day mass function (MF) which
may be different from the initial one. At the high-mass end, the two
will differ because of stellar evolution, but in a
stellar cluster there can also be differences between the IMF and
the present-day MF at the low-mass end due to effects of dynamical 
evolution. Thus, we will deliberately omit the I in the term \emph{IMF} 
throughout this paper.

  Even with the HST, direct observations of all but 
the brightest individual stars in extragalactic star clusters are generally 
impossible, so attempts to constrain the MF have to rely on indirect methods.
One approach which has gained popularity in recent years is to combine 
information about structural parameters for the clusters, typically derived 
from HST imaging, with 
velocity dispersion measurements obtained from integrated-light spectra
with high-dispersion spectrographs, and estimate the mass-to-light 
(M/L) ratios by application of the virial theorem. If the cluster ages
are known (for example from broad-band colors), the virial M/L estimates can 
then be compared with predictions by 
simple stellar population (SSP) models computed for various MFs. 
\citet{hf96a,hf96b} studied three young clusters
in the two nearby dwarf galaxies NGC~1569 and NGC~1705 and found masses of
several times $10^5 \, \msun$. Analyzing the same data and comparing
with SSP models, \citet{s98} found some evidence for variations in the MF 
slope between the three clusters.  Several other recent studies have
reported a mixture of ``normal'' (usually meaning 
Kroupa 2002 or 
Salpeter 1955-like) and top-heavy MFs \citep{sg01,men03,mar03,gg03,mgg03}.
While in principle straight-forward, the simple application of the virial 
theorem is in practice complicated by several factors, such as the assumption 
of virial equilibrium (which may not be valid for the youngest clusters) and
isotropy of the velocity distribution,
mass segregation (primordial or as a result of dynamical evolution),
presence of binary stars, and (especially for more distant systems)
crowding- and resolution effects. 
Evidence for a top-heavy MF has also been claimed based on Balmer line 
strengths measured on medium-resolution spectra for clusters in the peculiar 
galaxy NGC~1275 \citep{bro98}.

  In this paper we report new observations of four luminous young star
clusters in the two nearby irregular dwarf galaxies NGC~4214 and NGC~4449.  
HST photometry for the clusters has been published by 
\citet[][hereafter BHE02]{bhe02} and \citet[][GHG01]{ghg01}. 
A primary selection criterion for our spectroscopic follow-up was that the 
clusters be isolated and appear on as uniform a background as possible, 
allowing accurate measurements of their structural parameters and velocity
dispersions. In addition, the clusters all have estimated ages greater than 
100 Myrs.  With typical crossing times of $\sim10^6$ years the
assumption of virial equilibrium is expected to be a very plausible one for
these clusters.
We have obtained high-dispersion spectra with the HIRES spectrograph 
\citep{vogt94} on the Keck-I telescope and use these spectra to measure 
velocity dispersions  for the clusters. We have also observed a younger
cluster in the nearby spiral galaxy NGC~6946 with the NIRSPEC spectrograph 
on the Keck-II telescope. We have previously observed this cluster with 
HIRES \citep[hereafter LAR01]{lar01}, so the new data provide a welcome 
consistency check of our earlier results.  

  For the distances to NGC~4449 and NGC~6946 we assume 3.9 Mpc and 6.0 Mpc
\citep{ghg01,kar00}.  The distance to NGC~4214 has long been 
assumed to be around 4--6 Mpc and BHE02 used 4.8 Mpc. However, recent 
measurements by \citet{droz02} and \citet{maiz02} based on HST photometry of 
the red giant branch suggest significantly smaller values of $2.7\pm0.3$ Mpc 
and $2.94\pm0.18$ Mpc, respectively. Here we adopt a compromise of 
$2.8\pm0.3$ Mpc.  


\section{Observations and data reduction}

\subsection{WFPC2 imaging}

\begin{figure}
\begin{minipage}{16cm}
\plotone{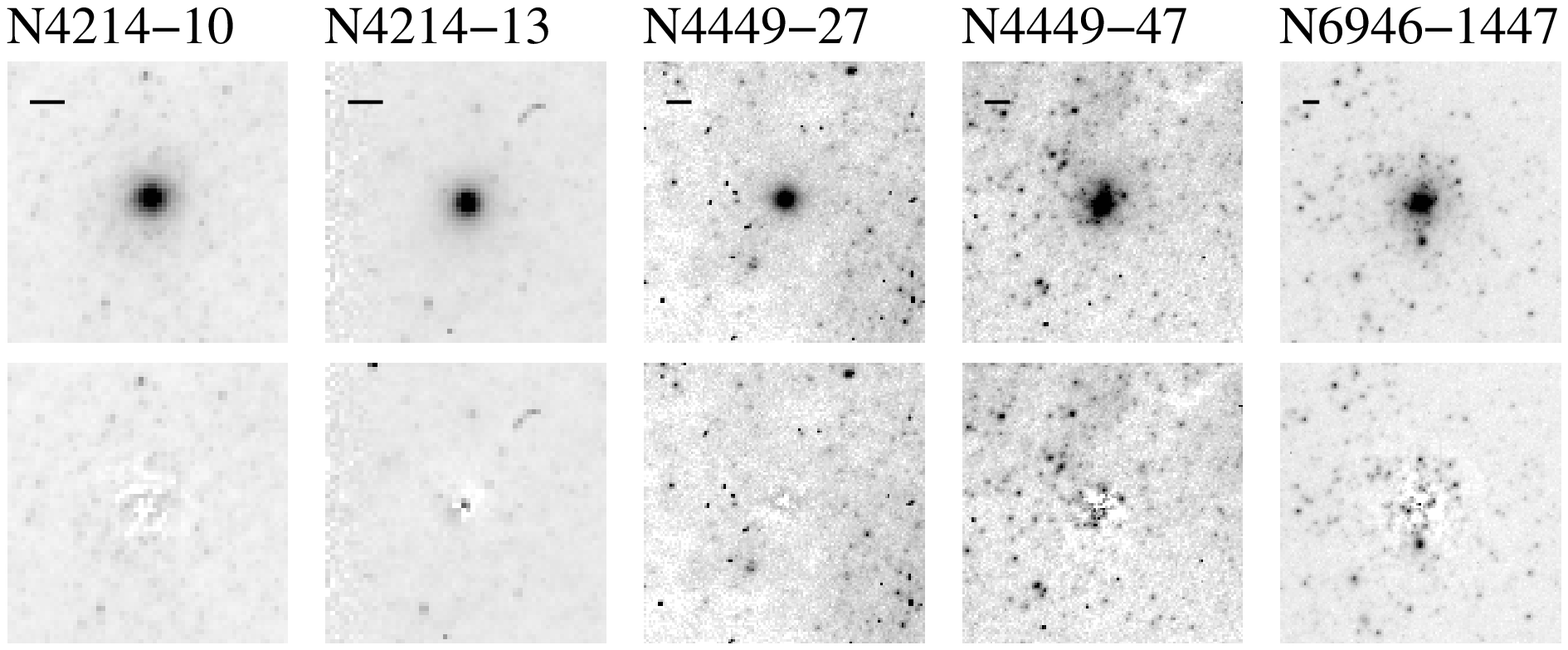}
\end{minipage}
\figcaption[larsen.f1.ps]{\label{fig:hstimg}HST F555W images of the five clusters.
Each top panel shows a $6\arcsec\times6\arcsec$ section of the image 
centered on the cluster. The bar in the upper left corner of each panel
indicates a linear scale of 10 pc.  The bottom panels show the residuals 
after subtraction of the best-fitting EFF model convolved with the
TinyTim PSF.
}
\end{figure}

  Table~\ref{tab:obsloghst} lists the cluster IDs and the HST datasets 
used for this analysis.  Coordinates for the clusters and discussion of 
their host galaxy properties can be found in GHG01, BHE02 and LAR01.  HST 
images of the clusters are shown in Fig.~\ref{fig:hstimg}.
When possible, multiple exposures were combined to reject cosmic 
ray hits, but in many cases we only had a single exposure in each band. The 
cluster N4214-13 was saturated on the F555W and longer F814W exposures 
obtained under programme 6569, so for this cluster we used only the F336W 
data and a single short F814W exposure from that programme.  Other images
in F606W, F555W and F814W were available for N4214-13 from programmes 
5446 and 6716 and these were all used for the analysis.  For the
two clusters in NGC~4449, only a single exposure in each of the F555W and 
F814W bands were used.  Exposures in F336W were also available for this galaxy 
from programme 6716 and were used by GHG01 for photometry, but had too low 
S/N for our size measurements.  For the photometry of clusters in NGC~4214, 
NGC~4449 and NGC~6946 we use the data published in BHE02, GHG01 and 
\citet{lar01}, and we refer to those papers for further details. The colors 
and $V$ magnitudes, corrected for Galactic foreground extinction, are listed 
in Table~\ref{tab:phot}. Note that GHG01 and BHE02 also included a correction
of $E(B-V)=0.05$ mag for internal reddening in the galaxies, which is
included for the photometry in Table~\ref{tab:phot}.

\subsection{Ages and reddenings}
\label{sec:age_red}

  Cluster ages and reddenings were determined by comparing Bruzual \& Charlot 
(2003, hereafter BC03) model predictions for the evolution of broad-band 
colors as a function of age for simple stellar populations (SSPs) with the 
observed cluster colors.  We used the BC03 models tabulated for a 
\citet{salp55} stellar MF extending down to $0.01\,\msun$, but the exact 
MF choice is not important
for the age determinations because the integrated colors of clusters are
dominated by stars in a narrow mass range near and just above
the turn-off. The BC03 models are also tabulated for a Chabrier MF, and
we verified that using this MF yielded nearly identical results to the 
Salpeter MF, as expected. The MF choice does of course have a significant 
impact on the \emph{mass-to-light} ratios, as will be discussed below.

\begin{figure}
\begin{minipage}{16cm}
\plotone{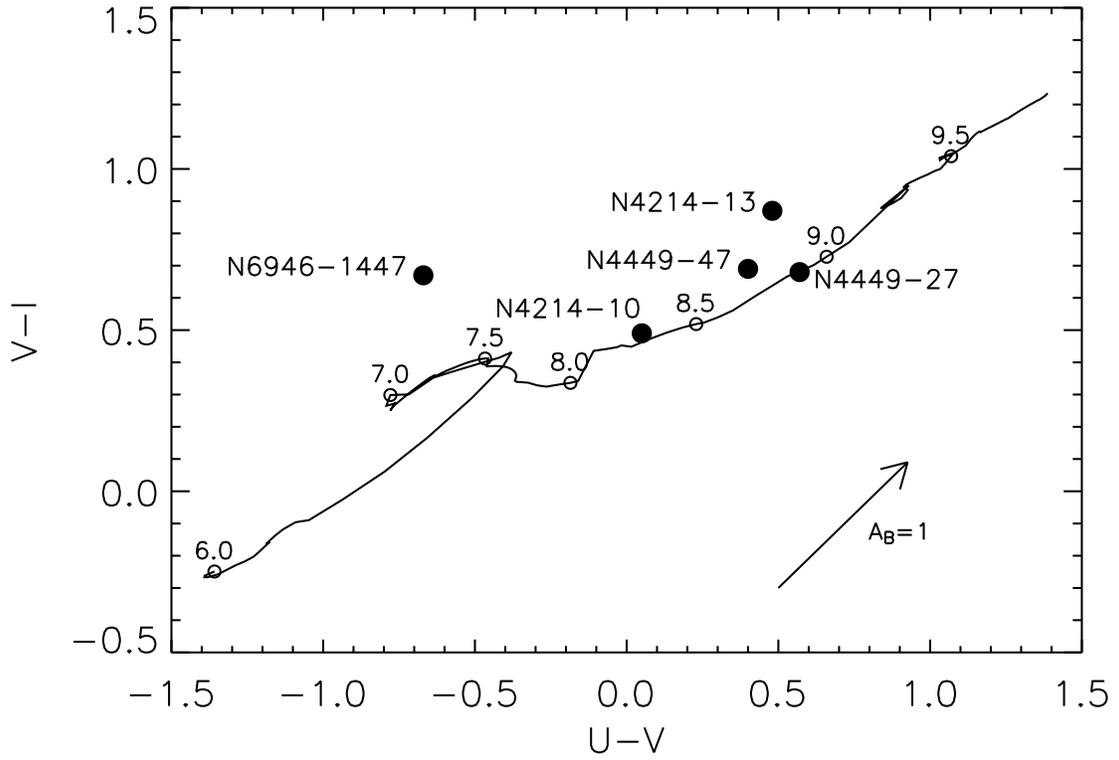}
\end{minipage}
\figcaption[larsen.f2.ps]{\label{fig:vi_uv}Two-color diagram showing Bruzual \& Charlot
(2003) models ($Z=0.008$) and observations for the clusters. The
logarithm of ages are indicated along the model sequence. The arrow 
indicates the reddening vector for $A_B=1.0$. Photometry is not
corrected for extinction internally in the galaxies (see Table~\ref{tab:phot})
}
\end{figure}

  Fig.~\ref{fig:vi_uv} shows the BC03 model sequence in the
($U-V$, $V-I$) plane for metallicity $Z=0.008$ and the 
five datapoints corresponding to the cluster photometry in 
Table~\ref{tab:phot} (without any additional reddening corrections).
Ages are indicated along the sequence.  The arrow indicates the reddening 
vector for $A_B=1$ mag, according to the Galactic extinction law in 
\citet{car89}. Ages and reddenings were determined for each cluster by 
projecting the observed colors along the reddening vector until the best 
match to the BC03 models was obtained.  We have only two colors for 4 of the 5 
clusters and for these objects the best match between model and observed 
colors simply corresponds to the intersection between a line drawn through 
the datapoints in Fig.~\ref{fig:vi_uv} along the reddening vector and the
model sequence.  For NGC6946-1447 more colors are available and an exact 
match to the model colors can not be achieved. For this cluster we selected 
the age/reddening combination corresponding to the closest match between the 
projection of the observed colors along the reddening vector and the 
BC03 models. 


  The best fitting age- and extinction estimates are given for $Z=0.004$,
$Z=0.008$ and $Z=0.02$ (solar) models in Table~\ref{tab:phot}.  While the 
metallicities
of the clusters themselves are unknown, estimates of oxygen abundances 
for H{\sc ii} regions in NGC~4214 and NGC~4449 are available in the
literature.  For two knots in NGC~4214, \citet{ks96} found 12 + $\log$(O/H) = 
$8.173^{+0.020}_{-0.022}$ and $8.267^{+0.015}_{-0.017}$, corresponding
to [O/H]$\sim-0.57$ (taking the Solar [O/H] from \citet{apf}). For 
NGC~4449, \citet{leq79} measured 12 $+$ $\log$(O/H) = 8.3 or [O/H]$=-0.49$. 
Assuming solar-scaled abundances, these Oxygen abundances correspond to 
$Z=0.0054$ and $Z=0.0065$, intermediate between the BC03 $Z=0.004$ and 
$Z=0.008$ models.  The [O/H] abundance in NGC~6946 was estimated for
two clusters near N6946-1447 by \citet{efre02}, who found 
12 $+$ $\log$(O/H) = 8.95$\pm$0.2, consistent with the Solar value.
Since the $Z=0.004$ models give a negative reddening for
N4449-27 (even accounting for the $E(B-V)=0.05$ internal reddening correction
implicit in the photometry) and the ages and reddenings derived for
NGC~4214 are nearly independent of the assumed metallicity, we will use the 
age and metallicity estimates derived for 
$Z=0.008$ for the clusters in NGC~4214 and NGC~4449 and those derived
for $Z=0.020$ for the cluster in NGC~6946.  The age estimates are then 
virtually identical to those tabulated by BHE02 and GHG01, who used 
Starburst99 models.  
For N6946-1447, it is interesting to note that the procedure adopted
here also yields a very similar age estimate to that obtained by LAR01,
where the \citet{gir95} ``S''-sequence calibration was used. While the
S-sequence method is attractive because of its empirical founding, it is
(at least for now) restricted to $UBV$ colors, and the multi-color
approach adopted here allows for a more robust estimate of the foreground
reddening.

  Obtaining realistic estimates of the errors on the derived ages and
reddenings is not straight-forward. The errors given in Table~\ref{tab:phot} 
are based on the photometric uncertainties only. They were estimated
by adding random offsets to the input photometry, drawn from Gaussian
distributions with standard deviations corresponding to the photometric
errors, and rederiving the ages and metallicities. This procedure was
repeated 100 times for each cluster, and the standard deviations of the
resulting distributions of age- and metallicity estimates were then taken
as estimates of the corresponding 1 $\sigma$ uncertainties. It is 
clear, however, that systematic uncertainties can dominate over the 
random errors. Table~\ref{tab:phot} illustrates the metallicity
dependence, but the models themselves are also uncertain. Stellar models
from different groups \citep[e.g.\ Padua vs.\ Geneva:][]{gir00,ls01}
differ significantly in their predictions
for the effective temperature, luminosity and lifetimes of red supergiants,
for example, which translates into uncertainties on the integrated colors
of simple stellar population models. To avoid artificially low
uncertainties on the log(age) estimates for the clusters with very small
errors on the photometry, we adopt 0.1 or the values in Table~\ref{tab:phot},
whichever is larger.

\subsection{Cluster sizes}

  For the measurements of cluster sizes we used the ISHAPE profile-fitting
code, which has been tested and described in \citet{lar99}.  Briefly, 
the code models the image of a cluster by assuming an analytic model for the
intrinsic profile and then convolves the model with a user-specified point 
spread function (PSF).  The shape parameters of the model are iteratively
adjusted until the best possible match to the data is obtained.  A 
particularly relevant feature for this work is that ISHAPE assigns weights to 
each pixel by measuring the standard deviation in concentric rings around
the center of the object. Pixels which deviate by more than 2$\sigma$
from other pixels in each annulus are assigned zero weight, 
effectively eliminating pixels affected by cosmic rays from the fitting
process.

We assumed cluster profiles of the form 
\begin{equation}
  P(r) \propto \left[1+(r/r_c)^2\right]^{-\alpha}
  \label{eq:eff}
\end{equation}
  These profiles differ from the classical \citet{king62} models 
for globular clusters in that they do not have a well-defined tidal radius,
but they have been shown by \citet{eff87} to provide very good fits to young
clusters in the Large Magellanic Cloud.  In the following we refer to them
as EFF profiles. During the fitting procedure, we allowed both the core 
radius $r_c$ and envelope slope parameter $\alpha$ to vary as free parameters.
The input WFPC2 PSFs were constructed using version 6.0 of the TinyTim 
software\footnote{TinyTim is available at
http://www.stsci.edu/software/tinytim/tinytim.html}
including a convolution with the ``diffusion kernel'' to simulate charge
diffusion between neighboring pixels.  Separate PSFs were generated for
each band at the position of each cluster. The residuals after
subtraction of the best-fitting EFF models, convolved with the TinyTim PSF,
are shown for the F555W images in the bottom panels of Fig.~\ref{fig:hstimg}.
The intensity scales in the bottom and top panels are identical.

  Table~\ref{tab:mfits} summarizes the EFF profile fits
for a fitting radius of $3\arcsec$. For each cluster we list the S/N
within the fitting radius, followed by the fitted FWHM (in arcsec,
corrected for the HST PSF) and envelope slope parameter $\alpha$.
The conversion from FWHM and $\alpha$ values to effective
(half-light) radii, needed for the virial mass determinations 
(Sec.~\ref{sec:results}), depends on the adopted outer radius of the 
cluster profiles. In the case of infinite cluster size and $\alpha>1$, 
the effective radius \reff\ is 
\begin{equation}  
  \reff \, = \, r_c \, \sqrt{(1/2)^{\frac{1}{1-\alpha}} -1}  
  \label{eq:reff1}
\end{equation}
where $r_c$ is the core radius in Eq.~(\ref{eq:eff}), related to the FWHM
as
\begin{equation}
  {\rm FWHM} \, = \, 2 \, r_c \, \sqrt{2^{1/\alpha}-1}.
\end{equation}
However, for $\alpha<1$ the total volume contained under the EFF profile
is infinite, and $\reff$ therefore undefined. If $\alpha$ is only slightly 
greater than 1, the $\reff$ computed from Eq.~(\ref{eq:reff1}) can be very 
large. Thus, more meaningful estimates of $\reff$ may be obtained by adopting 
a finite outer radius for the cluster profile:
\begin{equation}  
  \reff \, = \, r_c \left[\left\{\frac{1}{2}\left[
               \left(1+\frac{\rmax^2}{r_c^2}\right)^{1-\alpha}+1\right]
                \right\}^{\frac{1}{1-\alpha}}-1\right]^{1/2}
  \label{eq:reff2}
\end{equation}
where $\rmax$ is the adopted outer limit of the cluster. In 
Table~\ref{tab:mfits} we list $\reff$ values for $\rmax=3\arcsec$,
$5\arcsec$ and $\rmax=\infty$. While the \reff\ values do show
some dependence on the adopted \rmax, especially for clusters with
relatively shallow envelopes (N4214-13, N6946-1447), the difference
between $\rmax=3\arcsec$ and $\rmax=5\arcsec$
is comparable to the overall scatter in the measurements
on different images.  We also carried out fits within a fitting radius of 
$1\farcs5$, and found the results to be consistent with the numbers
in Table~\ref{tab:mfits} within the uncertainties. 

  We also carried out a series of King model fits to the cluster
profiles, the results of which are listed in Table~\ref{tab:kfits}.
The King profiles might seem a more obvious choice since they depend
on only two parameters, namely the core radius and concentration
parameter $c=r_t/r_c$, where $r_t$ is the tidal radius, and
there is no need to introduce an artificial cut-off radius. The
problem is that $r_t$ is often poorly constrained, leading to substantial 
uncertainties on \reff .  Nevertheless, the King profile fits in
Table~\ref{tab:kfits} are generally consistent with our estimates of the
half-light radii from the EFF fits. In the following we use the mean 
values listed in Table~\ref{tab:mfits} for fitting radius of 
$3\arcsec$ and $\rmax=3\arcsec$.

\subsection{HIRES and NIRSPEC spectroscopy}

  The four clusters in NGC~4214 and NGC~4449 were observed on May 8 and 
May 9, 2003, with the HIRES spectrograph on the Keck-I telescope. We used 
the C2 decker, providing a slit width of $0\farcs861$ and a resolution of 
$\lambda/\Delta\lambda=45000$.  A spectral range of 5450\AA --7800\AA\
was covered, with some gaps between the 20 echelle orders.  
Integration
times were about 3 hours for each object, typically split into 3 
exposures. This yielded signal-to-noise (S/N) ratios of 15--55 per pixel
in the dispersion direction in the combined spectra. The raw exposures
were reduced with the MAKEE package written by T.\ Barlow and tailored
specifically for HIRES spectra. MAKEE automatically performs bias 
subtraction and flat-fielding, then traces the echelle orders in 
the input images, extracts the spectra and performs wavelength calibration
using observations of calibration lamps mounted within the spectrograph
and night sky lines to fine-tune the zero-points.
Finally, the spectra were average combined using a sigma-clipping
algorithm to eliminate any remaining cosmic-ray hits not recognized by MAKEE. 

  The cluster in NGC~6946 was observed on Jul 13, 2002 with the NIRSPEC 
near-infrared spectrograph \citep{nirspecref} on Keck-II, using a 
$0\farcs432$ slit.  Observations were obtained for two echelle-mode
settings, covering the $H$ and $K$ bands with a spectral resolution 
of $\lambda/\Delta\lambda=25000$.  Integrations were made in pairs of 
240 s each, nodded by a few arcsec along the slit to allow sky 
subtraction. Observations of an A-type star (HR 44) were obtained for
removal of telluric lines from the spectra.  The data were reduced with the 
REDSPEC package written by L.\ Prato, S.\ S.\ Kim \& I.\ S.\ McLean,
but with two modifications by us: 1) we modified the extraction algorithm
to perform optimal extraction of the spectra \citep{horne86}, and 2) we 
implemented a
cross-correlation technique to fine-tune the wavelength scale of the
calibrator spectrum. Finally, the spectra were averaged in
the same way as for the HIRES data.

  Details of the HIRES and NIRSPEC observations are provided in 
Table~\ref{tab:obslogkeck}. In
addition to the cluster spectra, a number of template stars for the 
velocity dispersion measurements were also observed during both the 
HIRES and NIRSPEC runs. These are listed in Table~\ref{tab:tmpl_I_II} 
and \ref{tab:tmpl_III}. Absolute $M_V$ magnitudes are listed for each
star, based on \emph{Hipparcos} parallaxes \citep{per97} and the
$V$ magnitudes in the Bright Star Catalogue \citep{hl91}.
For the HIRES run we observed both giant
and supergiant template stars, since the clusters in NGC~4214 and
NGC~4449 are sufficiently old that giant stars might be more appropriate 
templates. For the NIRSPEC run, only the luminosity class (LC) I and II stars
were observed.

  Velocity dispersions were measured using the cross-correlation technique 
first described by \citet{td79} and applied to young clusters by
\citet{hf96a,hf96b}.  We have 
previously used this method in LAR01 for our analysis of N6946-1447 HIRES data 
and refer to that paper for details specific to our approach. 
In brief, the cluster spectra were cross-correlated with the spectra of the 
template stars, using the FXCOR task in the RV package within 
IRAF\footnote{IRAF is distributed by the National Optical Astronomical 
Observatories, which are operated by the Association of Universities for 
Research in Astronomy, Inc.~under contract with the National Science 
Foundation}.  
Prior to the cross-correlation, the spectra were
continuum-subtracted using a cubic spline and any remaining large
scale variations were eliminated by further applying a high-pass filter
cutting at a wavenumber of 10.
The FWHM of the peak of the resulting cross-correlation function (CCF) 
is a measure of the broadening of the stellar lines in the cluster spectrum, 
here assumed to be mainly due to the line-of-sight velocity dispersion, $v_x$.  
Because the spectra are continuum-subtracted and only the width of
the CCF peak is used, addition of a smooth continuum to the spectra (e.g.\ 
from early-type stars in the clusters) does not affect the velocity 
dispersion measurements.  In the original implementation of the method 
described by
\citet{td79}, the $v_x$ is determined as the quadrature difference between
the Gaussian dispersion of the CCF peak of the template versus object
spectra ($\mu$) and the internal velocity dispersion of the template,
$\tau$: 
\begin{equation}
  v_x^2 \, = \, \mu^2 \, - \, 2 \tau^2
  \label{eq:td79}
\end{equation}
Setting $v_x^2 = 0$ for the template spectrum, it is seen that $2\tau^2$ is 
simply the squared dispersion of the CCF peak for two template star spectra.
Thus, $v_x^2$ is the squared difference of the dispersions of the two
CCF peaks. A convenient feature of the cross-correlation method is that
instrumental resolution effects cancel out, as long as they are the same
for the object- and template spectra. In principle this also applies
to any intrinsic broadening of the stellar lines, for example by
macroturbulent motions in the atmospheres.

  If the CCFs do not have a Gaussian shape, it is unclear how to relate 
the FWHM values output by the FXCOR task to Gaussian sigmas. Thus, we 
adopted a slightly different approach than simply applying 
Eq.~(\ref{eq:td79}).  The relation between the FWHM of 
the CCF peak and $v_x$ was established by convolving the template star spectra 
with a series of Gaussians corresponding to $v_x$ values
bracketing the values expected for the clusters, and then cross-correlating 
the broadened template star spectra with the unbroadened ones. Thus, each 
cross-correlation product of broadened and un-broadened spectra led to an 
estimate of the CCF peak FWHM for the corresponding broadening, which could
then be compared with the FWHM of the cluster versus template CCF.  The 
cross-correlation method 
does not rely on any individual, strong lines, but utilizes the multitude of 
fainter lines from late-type stars that are present in the spectra.  In
fact, it is better to avoid strong lines like the Ca II triplet
where saturation effects set in. 
For this work, we performed the analysis separately for each echelle order, 
though not all echelle orders were included. Some orders included only a 
few suitable lines, while others were contaminated by sky lines. 
For the HIRES spectra we used 8 out of the 20 available echelle orders 
while 6 out of the 15 echelle orders 
were used for the NIRSPEC data.

\begin{figure}
\begin{minipage}{14cm}
\plotone{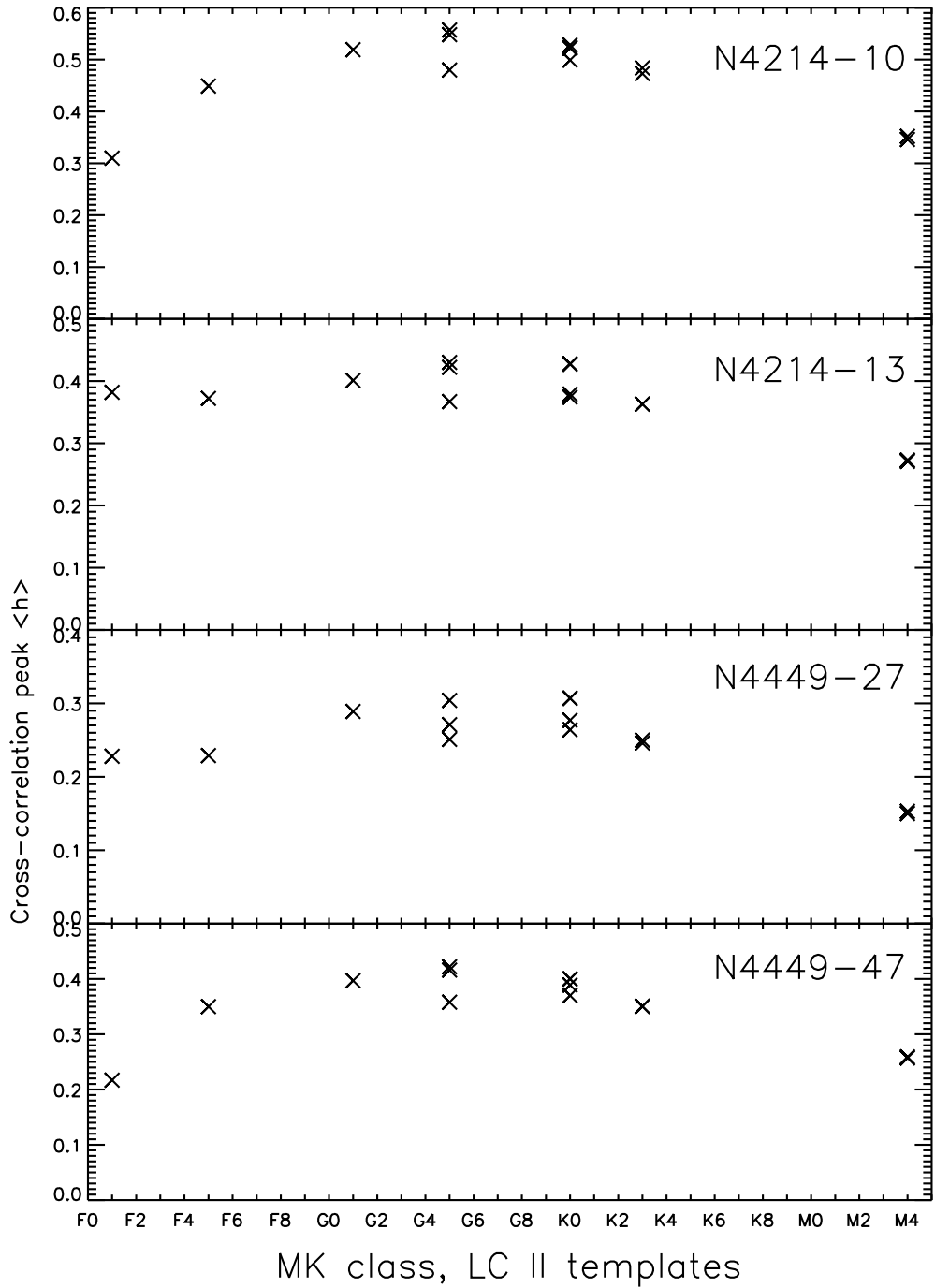}
\end{minipage}
\figcaption[larsen.f3.ps]{\label{fig:mk_h}Cross-correlation peak amplitude vs.\
  spectral type of the template star for luminosity class II bright giants.
}
\end{figure}

  Figure~\ref{fig:mk_h} shows the mean peak amplitude of the CCF as a function 
of template star spectral type for the HIRES data, averaged over all echelle 
orders included in the analysis.  Only stars of luminosity class II are 
included in this figure, but similar plots were made and inspected for the 
LC I and III templates.  For all four clusters, the maximum is reached for 
late G and early K-type, indicating that the cluster spectra are best
matched by such stars.  Thus, we discard stars earlier than G0 
and later than M0.

  Most of the uncertainties in the cross-correlation analysis are probably 
systematic rather than random. Part of the uncertainty is due to the
many different parameters involved in the reduction, such as the details of
the continuum-subtraction, how the filtering of the spectra is done,
and which wavelength range is used for the cross-correlation. We found that
the derived velocity dispersions could vary by 0.5--1 km/s depending on the 
parameter settings. In particular, Tonry \& Davis applied both a high-pass 
filter (to remove large-scale variations in the spectra) and low-pass filter 
(to suppress noise) in their analysis, while we only use a high-pass filter. 
For the low-pass filter we found our velocity dispersions to be rather 
sensitive to the exact bandpass of the filter used, though there was a general 
trend for $v_x$ to increase when a filter was applied.  Also, contrary to the 
high-pass filtering, low-pass filtering of the data actually led to an 
increase in the scatter of the $v_x$ values. Therefore we decided not to 
apply a low-pass filter, but note that the $v_x$ values could be somewhat
underestimated because of this. 

Another potential problem is that 
the atmospheres of the template stars may have different macroturbulent 
velocities than the cluster stars, as the macroturbulent velocities in 
red giants and supergiants are known to depend on luminosity class 
\citep{gt87}. Because the velocity dispersions in the clusters are 
comparable to, or in some cases smaller than, the macroturbulent velocities 
in the atmospheres of the template stars ($\sim10$ km/s), this could 
potentially lead to serious systematic errors in the cluster velocity 
dispersions.  Table~\ref{tab:vdmeas} lists the 
$v_x$ values derived for the clusters using template stars of LC
I, II and III.  
The measurements show a clear trend of increasing $v_x$ versus luminosity
class of the template star, consistent with a decrease in macroturbulent
broadening of the stellar lines for less luminous giants.
In order to determine which stars might provide the most suitable templates 
for our clusters, we compared the expected luminosities of the brightest
late-type stars of a given age, using stellar isochrones from the Padua 
group \citep{gir00}, with the tabulations of absolute $M_V$ magnitudes versus 
luminosity class in \citet{sk82}.  
For ages of $10^8$ years, $5\times10^8$ years and $10^9$ years, the brightest 
red giants have $M_V=-4.5$, $-3.0$ and $-2.5$, respectively, while \citet{sk82} 
tabulates typical $M_V$ magnitudes of $-2.3$ and $+0.7$ for K-type stars 
of LC II and III.  The $M_V$ magnitudes given by Schmidt-Kaler generally
agree quite well with the values listed in 
Table~\ref{tab:tmpl_I_II}--\ref{tab:tmpl_III}. This comparison suggests
that LC II bright giants might be the more suitable templates. 



\begin{figure}
\begin{minipage}{14cm}
\plotone{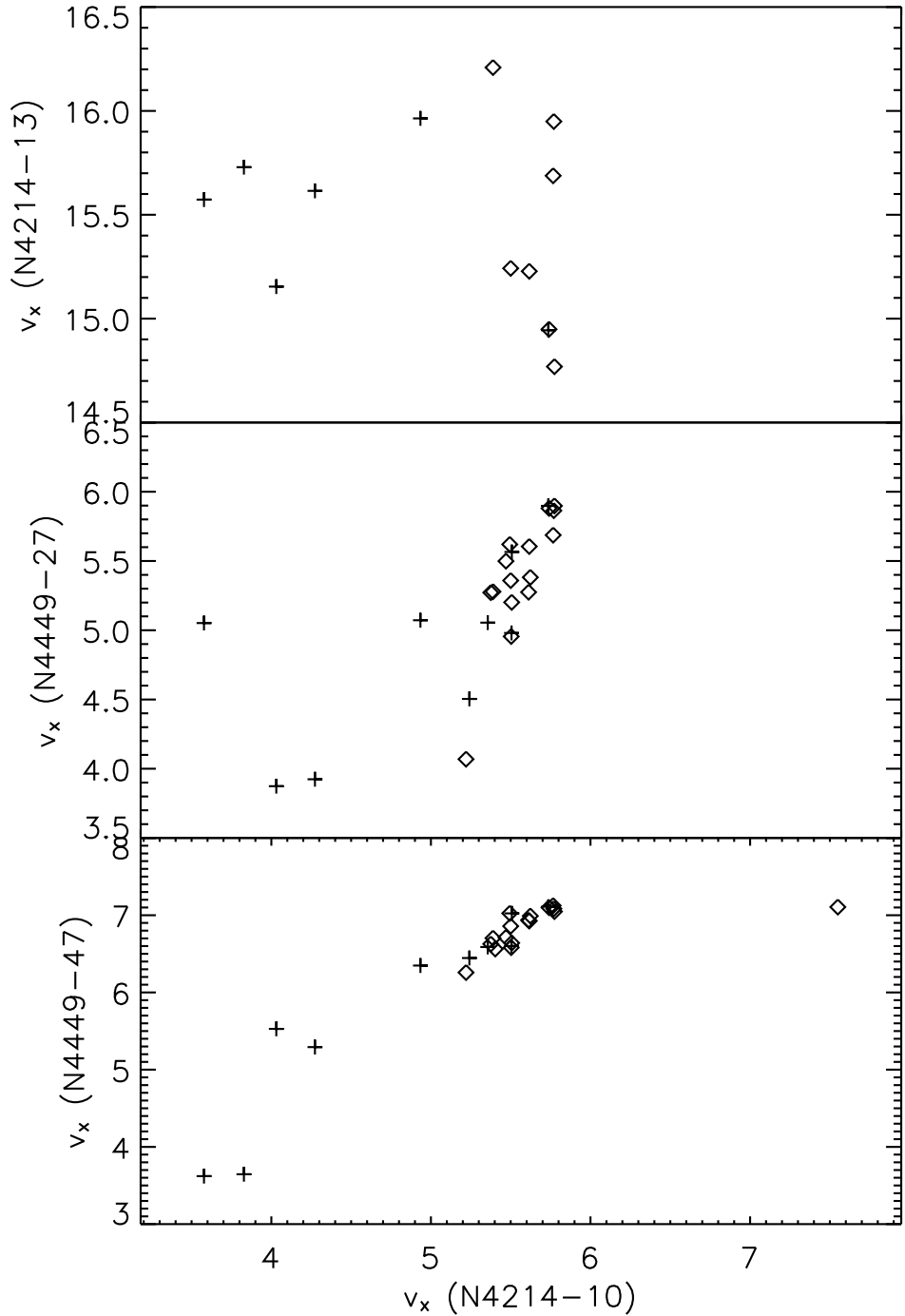}
\end{minipage}
\figcaption[larsen.f4.ps]{\label{fig:vdcmp}Comparison of velocity dispersion 
  measurements using different template stars. Each datapoint
  represents measurements based on one template star.
  Measurements for template stars of luminosity classes II and III
  are shown with plus markers and diamonds, respectively.
}
\end{figure}

\begin{figure}
\begin{minipage}{14cm}
\plotone{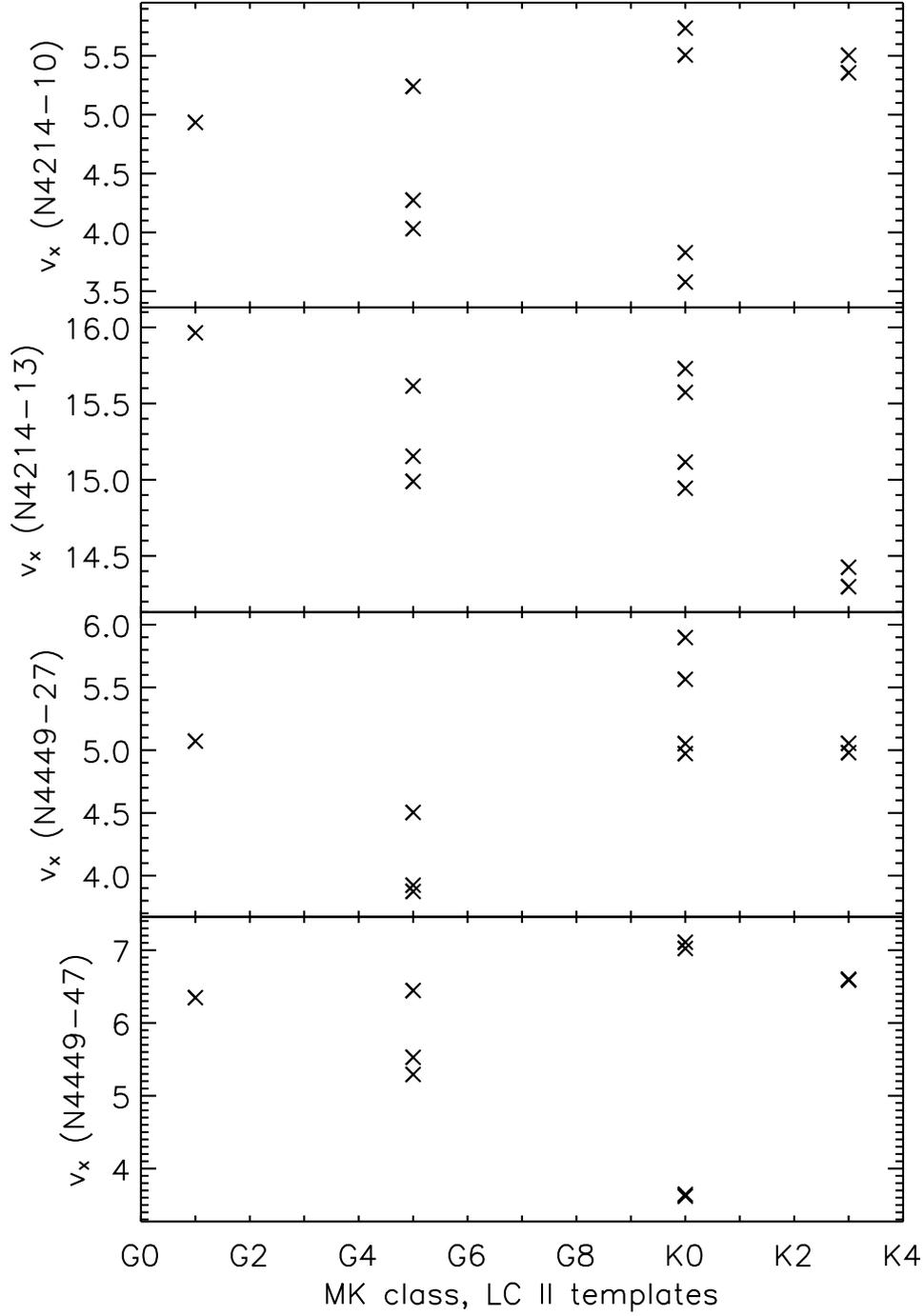}
\end{minipage}
\figcaption[larsen.f5.ps]{\label{fig:vd_mk}Velocity dispersion versus
  spectral type of the template star.
}
\end{figure}

  To further test how the choice of a particular template star might affect 
the velocity dispersions, we derived $v_x$ values for each cluster using each 
template star individually, i.e.\ the relation between FWHM of the CCF and 
$v_x$ was estimated by broadening each template with Gaussians and then
cross-correlating the broadened spectra with the unbroadened spectra of the
same stars.  In Fig.~\ref{fig:vdcmp} we plot these individual $v_x$ 
measurements for each cluster, based on each LC II and III template 
star, versus the corresponding $v_x$ for the cluster N4214-10 (somewhat 
arbitrarily chosen as a reference).  Measurements based on LC II and LC III 
stars are shown with plus ($+$) markers and diamonds ($\diamond$).  We 
have excluded measurements with a scatter of more than 3 km/s between the 
echelle orders.  If the derived $v_x$ values were uncorrelated with 
the choice of template star, we would expect only a random scatter
in Fig.~\ref{fig:vdcmp}.  Instead, the $v_x$ measurements clearly depend 
systematically on the choice of template star. Depending on the template,
the $v_x$ values may differ by as much as
3--4 km/s. There are clear systematics depending on whether the LC II 
and LC III templates are used, but some scatter even within the same
luminosity classes. We also tested if the $v_x$ values depend on the 
spectral type
of the template star. Figure~\ref{fig:vd_mk} shows the $v_x$ values
versus spectral type for the LC II bright giants.  Here, no clear correlation 
is seen, suggesting that a good match between the temperature of the template 
star and stars in the cluster is less critical.

  Presumably the best approach would be to select template stars with
very similar surface gravities to those in the cluster. However, such
detailed information is rarely available, and 
we simply adopt the mean of the LC II and III values
in Table~\ref{tab:vdmeas} for the clusters in NGC~4214 and NGC~4449. 
For the cluster in NGC~6946, which is much younger, we use template
stars of luminosity classes I and II.  We assign the same error bar 
of 1.0 km/s to all measurements, bracketing both of the values based
on LC II and LC III stars for the clusters in NGC~4214 and NGC~4449.

  In LAR01 we derived a velocity disperson of 10.0 km/s for N6946-1447,
based on HIRES data. This is somewhat higher than the 8.8 km/s obtained
from the NIRSPEC data, but it is worth noting that a value of 9.4 km/s was 
derived from the HIRES 
data if only the three best-fitting templates were used. It is not
clear \emph{a priori} that the HIRES and NIRSPEC data sample the same
stars, and effects such as mass segregation might be responsible (at
least partly) for the different velocity dispersions derived from the
two datasets. Arguing against such an effect is the fact that the red
supergiants span only a narrow mass range: according to isochrones by 
\citet{gir00}, stars above the main sequence turn-off have
masses between 16.3\msun\ and 17.1\msun\ for log(age)=7.10.  We do not
know to what extent N6946-1447 might be mass segregated. From 
Tables~\ref{tab:mfits} and \ref{tab:kfits}, the size tends to decrease 
for observations made at longer wavelengths (this was also noted by LAR01),
as if the red supergiants are predominantly concentrated near the center.
This suggests that some mass segregation may be present, although other
effects (e.g.\ differential reddening) could also play a role.  At any rate, 
the comparison of velocity dispersion measurements obtained with the two 
different spectrographs, covering very different wavelength ranges and 
with different instrumental resolutions, provides another estimate of 
the uncertainties and suggests that our 1 km/s estimated errors are 
reasonable.

\section{Results and discussion}
\label{sec:results}


  Assuming that the clusters are in virial equilibrium, the
line-of-sight velocity dispersion $v_x$, cluster half-light radius 
\reff\ and virial mass \mvir\ are related as
\begin{equation}
  M_\mathrm{vir} \, \approx \, 9.75 \frac{\reff \, v_x^2}{G}.
  \label{eq:vir2}
\end{equation}
\citep[][p.\ 11--12]{spit87}.  The typical crossing times are of order
$\reff / v_x \sim 1$ Myr, so the clusters studied here are many crossing 
times old and the assumption of virial equilibrium appears fully justified 
(possibly with the exception of N6946-1447, which is only $\sim10$
crossing times old).  The factor in front of Eq.~(\ref{eq:vir2}) depends on 
the density profile of the cluster and on the assumption of velocity 
isotropy, but the value used here should be accurate to $\sim10\%$ for 
most realistic cluster profiles \citep{spit87}. The simple relation
given by Eq.~(\ref{eq:vir2}) hides a number of uncertainties, such as
the assumption of isotropy, contribution to line broadening by binary 
stars, and mass segregation, which are difficult to quantify. 

Table~\ref{tab:ppar} summarizes the observed and derived parameters for 
the five clusters. 
All five clusters have masses in 
excess of $10^5\msun$, and two of them are more massive than $10^6\msun$. 
We also tabulate $V$-band mass-to-light ratios ($M/L_V$),
based on the virial mass estimates and the integrated $M_V$ magnitudes 
corrected for reddening.  
The central surface brightness $\mu_0$ (in mag arcsec$^{-2}$) 
and integrated $V$ magnitude within radius $r$ of EFF profiles are related as
\begin{equation}
  10^{-0.4\,\mu_0} \, = \,
    \frac{(1-\alpha) 10^{-0.4\,V(r)}}
         {\pi r_c^2 \left[ (1+r^2/r_c^2)^{1-\alpha}-1\right]}
\end{equation}
for $\alpha\neq1$ and with $r$ and $r_c$ measured in arcsec.
In order to compute the central densities $\rho_0$ we use the relation
\begin{equation}
  \rho_0 = \frac{3.44\times10^{10}}{P r_c} 10^{-0.4\mu_0} 
    \left(\frac{M}{L}\right) M_{\odot} \, \rm{pc}^{-3},
\end{equation}
with $P\approx2$ and $r_c$ in parsecs \citep{pk75,wb79}.  We use
the dynamical estimates of the $M/L$ ratios, but note that the errors 
on $\rho_0$ do not include the uncertainty on $M/L$.  In any case,
the $\rho_0$ values should be considered order-of-magnitude estimates only,
as the cores of the clusters are not well resolved. The mean densities 
within the half-mass radius, which are somewhat more robust, are 
listed as $\rho_{\rm hmr}$. Note that for most realistic cluster profiles, 
the 3-dimensional half-mass radius is larger than the 2-dimensional 
(projected) effective radius by about a factor 4/3 \citep{spit87}. This
correction factor was included in the computation of the 
$\rho_{\rm hmr}$ values.

  For N6946-1447, the analysis is complicated by the fact that this
cluster has a very extended shallow envelope and no clear outer boundary,
as discussed in LAR01.  Therefore the half-light radius is also uncertain. 
In LAR01 we used aperture photometry in concentric apertures to obtain 
an estimate of the half-light radius, but the dynamical mass itself was
derived by modelling the density profile of the cluster assuming
hydrostatic equilibrium. This approach yielded a somewhat smaller
estimate of the dynamical mass than a direct application of the
virial theorem, but had its own uncertainties in that the WFPC2 PSF was
not taken into account in the profile modeling and the boundary
conditions for pressure and density are unknown. For a given value
of the velocity dispersion, the mass derived by this method was lower
by about 25\% than the valued obtained by Eq.~(\ref{eq:vir2}) using
the structural parameters in Table~\ref{tab:ppar}. This difference 
is not strongly significant given the 16\% uncertainty on the \reff\
estimate, which translates to a similar uncertainty on the mass.

\begin{figure}
\begin{minipage}{16cm}
\plotone{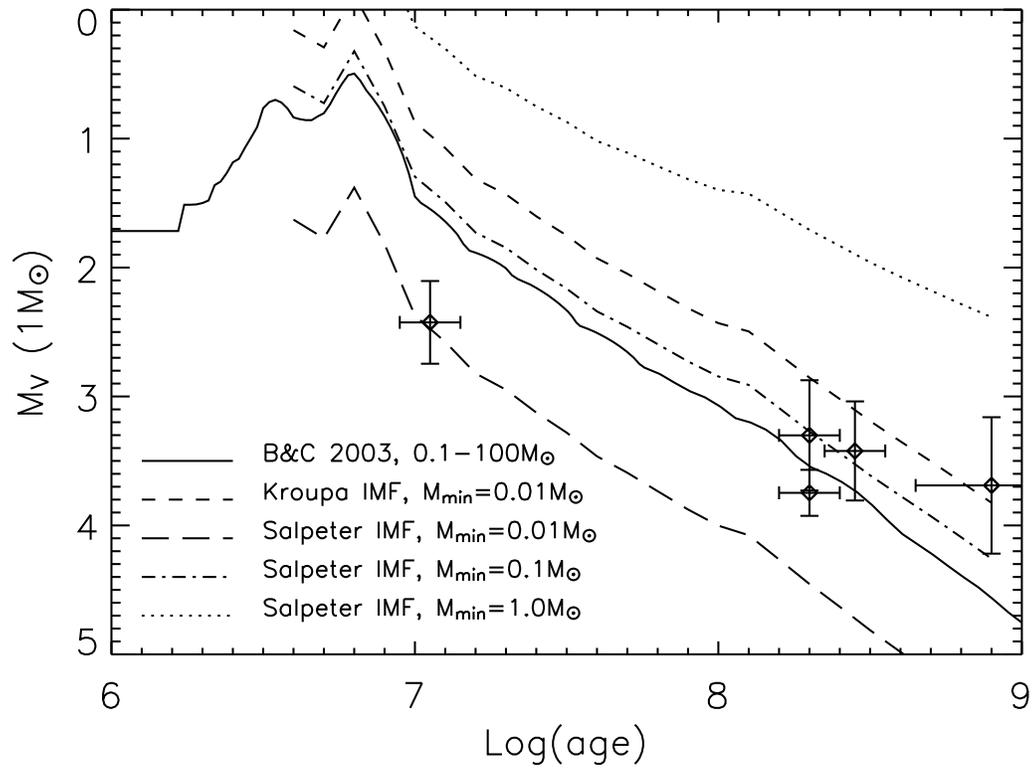}
\end{minipage}
\figcaption[larsen.f6.ps]{\label{fig:pmtol}Comparison of observed 
  mass-to-light ratios for young stellar clusters in NGC~4214, NGC~4449 
  and NGC~6946 with SSP models calculated for various stellar mass functions.
}
\end{figure}

  In Fig.~\ref{fig:pmtol} we finally compare the observed mass-to-light 
ratios (here expressed as $M_V$ magnitude per solar mass) with SSP models 
computed for a variety of stellar MFs. The solid curve shows Bruzual \& 
Charlot (2003) models for a Salpeter law with a lower mass limit of 
$0.1\msun$. The remaining curves were calculated by us, populating stellar 
isochrones from \citet{gir00} according to Salpeter laws with lower
mass limits of $0.01\msun$, $0.10\msun$ and $1.0\msun$ (long-dashed, 
dotted-dashed and triple-dotted-dashed curves) and a Kroupa MF (short-dashed 
curve).  Our model sequence for a Salpeter law extending to $0.1\msun$
is brighter than the Bruzual \& Charlot models by up to $\sim0.3$ mag,
with the difference increasing at higher ages. This might be partly due to
different approaches in the treatment of dark remnants, which are ignored
in our calculations.
All curves are shown for $Z=0.008$, but using $Z=0.02$ would shift 
them downwards by only 0.1-0.2 mag. Thus, it appears reasonable to plot 
N6946-1447 on the same graph, even if its metallicity may be somewhat 
higher than for the other clusters.  

Based on Figure~\ref{fig:pmtol}, at least the four older clusters appear
consistent with a Salpeter-type MF truncated at $0.1\msun$ or a Kroupa-type 
MF, while N6946-1447 may be somewhat bottom-heavy. The results for this
youngest object, however, may be more uncertain due to its young age
and uncertain size.
The
current data and many uncertainties inherent in the analysis do not allow
us to distinguish between the Salpeter and Kroupa IMFs, but any MF with a
significant excess or deficiency of low-mass stars, such as Salpeter MFs 
truncated at $1.0\msun$ or extending all the way down to $0.01\msun$, 
appears unlikely. Formally, the M/L ratio for NGC4214-13 appears to
be too high for a Kroupa-type MF, although the offset is reduced
if we shift the Kroupa curve downwards by an amount corresponding to
the difference between our Salpeter ($M_{\rm min}=0.1\msun$) curve and 
the Bruzual \& Charlot model.  While some claims for top-heavy MFs have been 
made by various authors (e.g.\ \citet{sg01}), inclusion of a Salpeter MF 
extending down to $0.01\msun$ in Fig.~\ref{fig:pmtol} is perhaps mostly a 
numerical exercise, but does serve to illustrate at what level the MF is 
actually 
constrained.  

\begin{figure}
\begin{minipage}{16cm}
\plotone{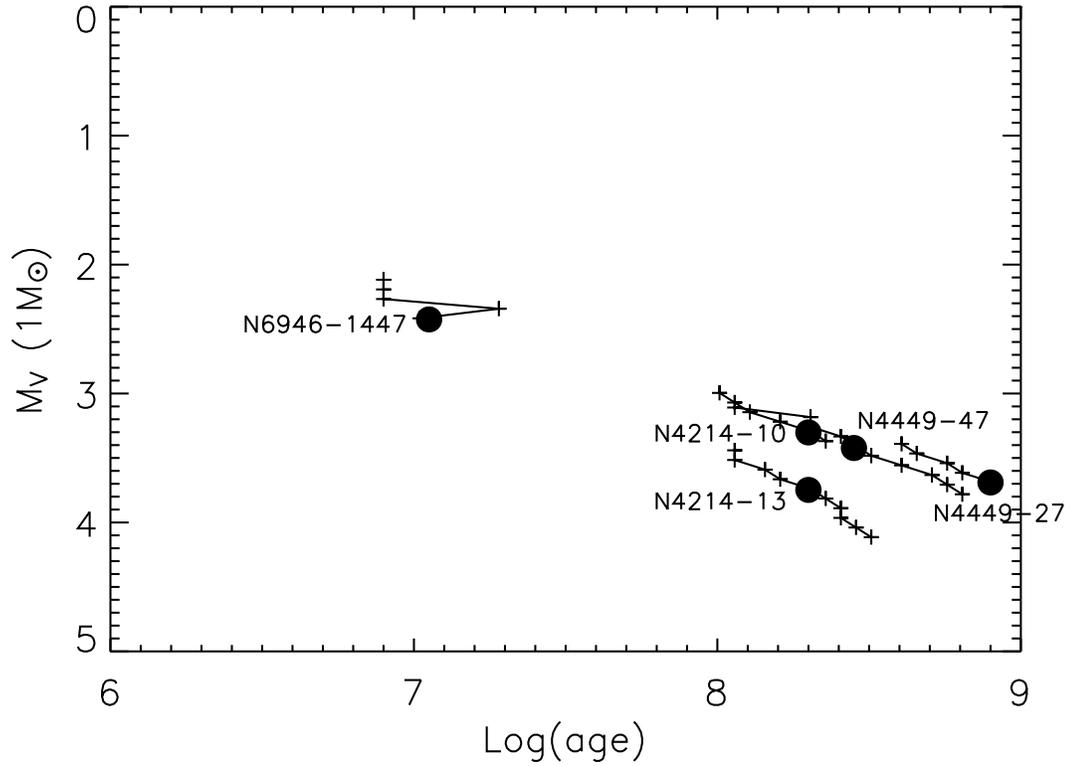}
\end{minipage}
\figcaption[larsen.f7.ps]{\label{fig:pab_age}Change in ages and
  mass-to-light ratios for different extinction values.
  The filled circles indicate the
  locations of the clusters in the log(age) vs.\ $M_V(1 \msun)$ 
  plane for the best-fitting extinction values, while the curves
  through each datapoint indicate the effect of varying the extinction
  within a range of $\pm0.5$ mag (but not less than $A_B=0.0$) with 
  respect to the $A_B$ values in Table~\ref{tab:ppar}.
}
\end{figure}

  In Fig.~\ref{fig:pab_age} we plot curves illustrating the range of ages 
and mass-to-light ratios obtained for each cluster by keeping the extinction 
correction fixed at a range of values between $\pm0.5$ mag with respect
to the $A_B$ values in Table~\ref{tab:ppar} (but avoiding negative
extinctions).  The filled circles show the best-fitting values from 
Table~\ref{tab:ppar}, obtained by allowing the 
extinction to vary as a free parameter, as in Fig.~\ref{fig:pmtol}.
Increasing the extinction correction 
makes the clusters appear intrinsically more luminous, as well as bluer and 
thus generally younger. For most of the clusters, the "reddening vector" in 
Fig.~\ref{fig:pab_age} happens to be roughly parallel to the model tracks
in Fig.~\ref{fig:pmtol}, so 
that even fairly large uncertainties on the extinction are not expected to 
strongly affect the comparison of M/L ratios with the SSP models.
The exception is NGC6946-1447, which is in an evolutionary phase dominated 
by rapid color variations due to the appearance of red supergiants. 
Therefore, the age does not vary smoothly with the assumed extinction as
for the other clusters.

  The $M/L_V$ ratios are inversely proportional to the assumed distance so 
the comparison in Fig.~\ref{fig:pmtol} is not strongly affected by small 
uncertainties in the distances.  If the distance to 
NGC~4214 is as large as previously assumed, i.e.\ about 5 Mpc, the two 
datapoints for this galaxy would shift upwards by 0.6 mag in 
Fig.~\ref{fig:pmtol}. The 
datapoint for NGC4214-10 would then fall about 0.2 mag above the model curve 
calculated for the Kroupa MF, but would still be consistent with it within 
the uncertainties. NGC4214-13 would fall about 0.3 mag below the Kroupa curve.
We note, however, that the present estimate of the distance of 2.8 Mpc, which 
is based on photometry on resolved stars, is probably more accurate than
the old value base on Hubble flow.


  The masses, sizes and central densities of the five clusters studied
here are well within the range spanned by Milky Way globular clusters 
(GCs).  Using GC luminosities tabulated in \citet{har96} and assuming a
$M/L_V$ ratio of $1.5$ \citep{mac00}, the median mass of Milky Way
GCs is $1.0\times10^5\msun$. All of the clusters in Table~\ref{tab:ppar}
have masses greater than $10^5 \, \msun$ and two of them
(N4214-13 and N6946-1447) have masses similar to even the most massive of 
Milky Way GCs, such as $\omega$ Cen. 
The physical dimensions (half-light radii) of these young clusters 
are also very similar to those of old globular clusters.
Curiously, the sizes of star clusters show little or no correlation with
mass \citep{zepf99,lar04}, and in this regard it is interesting to note that 
the most \emph{compact} of the five clusters analysed here (N4214-13) is also 
the second-most \emph{massive}.
While the cluster in NGC~6946 is only $\sim15$ Myrs old, the four clusters 
in NGC~4214 and NGC~4449 already have ages of several hundred Myrs, and it 
appears likely that they will eventually evolve into objects that are
very similar to the globular clusters presently observed in the Milky
Way.  
It is also worth noting that these clusters are in an age range where
the near-infrared light is expected to be dominated by asymptotic
giant branch stars \citep{lan99,ml02}, and it would be interesting
to obtain infrared spectra to look for AGB features.

\section{Summary and conclusions}

  We have measured velocity dispersions and structural parameters for
five luminous young star clusters (``super-star clusters'') in the
nearby spiral galaxy NGC~6946 and in two irregular dwarf galaxies,
NGC~4214 and NGC~4449. The five clusters are all well resolved on
HST/WFPC2 images, allowing us to accurately measure their structural
parameters and obtain spectra free of contamination.  The dynamically
derived mass-to-light ratios are at least as high as those predicted by 
SSP models for a Kroupa-type MF or a Salpeter law extending
down to $0.1\msun$. Within the uncertainties, we cannot distinguish
between these two possibilities, but the data are inconsistent 
with a MF with a significant deficiency of low-mass stars relative
to either, such as a Salpeter MF truncated at $1\msun$. The cluster in 
NGC~6946 has previously been observed with the HIRES spectrograph on Keck-I,
while the data used here were obtained with the NIRSPEC spectrograph
on Keck-II. The velocity dispersion derived from the NIRSPEC data
is consistent with our previous estimate from HIRES data within
about 1 km/s.  The masses, central densities and sizes of the five
clusters are within the same range spanned by Milky Way globular
clusters, and it is difficult to point to any differences between
these young star clusters and ``classical'' GCs other than age.

\acknowledgments

  JPB and SSL acknowledge support by National Science Foundation grant 
AST-0206139 and HST archival grant AR-09523-01-A. DAH acknowledges
National Science Foundation grant AST-0204922. We are grateful to the
anonymous referee for a number of constructive comments which helped
improve the paper.

\onecolumn

\begin{deluxetable}{lrrr}
\tablecaption{\label{tab:obsloghst}Observation log for HST imaging
  observations of the clusters. N4214-13 was saturated in the
  longer F555W and F814W exposures obtained under program 6569.}
\tablehead{Object & Program ID & Filter & Exposure}
\startdata
N4214-10  & 6569 & F336W & $260 + 2\times900$ s \\
          &      & F555W & $2\times600$ s       \\
	  &      & F814W & $100 + 2\times600$ s \\
N4214-13  & 5446 & F606W & 80 s \\
          & 6569 & F336W & $260 + 2\times900$ s \\
	  &      & F814W & 100 s \\
	  & 6716 & F555W & 200 s \\
	  &      & F814W & 200 s \\
N4449-27  & 6716 & F555W & 200 s \\
          &      & F814W & 200 s \\
N4449-47  & 6716 & F555W & 200 s \\
          &      & F814W & 200 s \\
N6946-1447 & 8715 & F439W & $2\times1100$ s \\
           & 8715 & F555W & $2\times300$ s \\
           & 8715 & F814W & $2\times700$ s \\
\enddata
\end{deluxetable}

\begin{deluxetable}{lrrrrr}
\tablecaption{\label{tab:phot}Photometric parameters for the clusters, 
 taken from BHE02 and GHG01. Photometry corrected for Galactic foreground
 reddening only.}
\tablehead{ & N4214-10       & N4214-13       & N4449-27      & N4449-47 & N6946-1447}
\startdata
$V_0$      & $17.08\pm0.01$  & $16.37\pm0.00$ & $18.35\pm0.01$& $17.58\pm0.01$ &  $15.70\pm0.01$ \\
$(U-V)_0$   & $0.05\pm0.01$  & $0.48\pm0.01$  & $0.57\pm0.05$ & $0.40\pm0.02$   &  $-0.67\pm0.01$ \\
$(U-B)_0$   &       -        &       -        &       -       &       -         &  $-0.70\pm0.01$ \\
$(B-V)_0$   &       -        &       -        &       -       &       -         &  $0.03\pm0.01$ \\
$(V-I)_0$   & $0.49\pm0.01$  & $0.87\pm0.01$  & $0.68\pm0.01$ & $0.69\pm0.01$   &  $0.67\pm0.01$  \\
$Z=0.004$: \\
\ Log(age)    & $8.30\pm0.04$  &  $8.35\pm0.04$ & $9.35\pm0.33$ & $8.65\pm0.06$ &   $7.02\pm0.01$  \\
\ $A_B$ (mag) & $0.03\pm0.08$  &  $0.91\pm0.10$ & $-0.52\pm0.32$ & $0.30\pm0.07$ &  $0.45\pm0.02$  \\
$Z=0.008$: \\
\ Log(age)    & $8.30\pm0.03$  &  $8.30\pm0.03$ & $8.90\pm0.25$ & $8.45\pm0.05$ &   $7.00\pm0.01$  \\
\ $A_B$ (mag) & $0.09\pm0.04$  &  $1.09\pm0.05$ & $-0.01\pm0.37$ & $0.48\pm0.05$ &  $0.72\pm0.02$  \\
$Z=0.020$: \\
\ Log(age)    & $8.25\pm0.03$  &  $8.25\pm0.03$ & $8.61\pm0.11$ & $8.40\pm0.03$ &   $7.02\pm0.09$  \\
\ $A_B$ (mag) & $0.15\pm0.02$  &  $1.14\pm0.03$ & $0.37\pm0.17$ & $0.51\pm0.06$ &   $-0.06\pm0.03$ \\
\enddata
\end{deluxetable}

\begin{deluxetable}{lrrrrrr}
\tablecaption{\label{tab:mfits}Size measurements for the clusters based
  on EFF fits within a $3\arcsec$ fitting radius.  }
\tabletypesize{\small}
\tablehead{  & S/N & FWHM & $\alpha$ & \multicolumn{3}{c}{$\reff$} \\
             &     &      &      & $\rmax=3\arcsec$ & $\rmax=5\arcsec$ & $\rmax=\infty$}
\startdata
N4214-10 \\
\ F336W (WF3) & 290 & $0\farcs215$ & 1.23 & $0\farcs333$ & $0\farcs368$ & $0\farcs550$ \\
\ F555W (WF3) & 980 & $0\farcs231$ & 1.28 & $0\farcs326$ & $0\farcs353$ & $0\farcs458$ \\
\ F814W (WF3) & 750 & $0\farcs265$ & 1.43 & $0\farcs299$ & $0\farcs312$ & $0\farcs338$ \\
\ Mean        & & & & $0\farcs319\pm0\farcs010$ & $0\farcs344\pm0\farcs017$ & $0\farcs449\pm0\farcs070$ \\
N4214-13 \\
\ F606W (PC)  & 340 & $0\farcs099$ & 1.14 & $0\farcs217$ & $0\farcs251$ & $0\farcs626$ \\
\ F336W (WF3) & 280 & $0\farcs013$ & 0.99 & $0\farcs152$ & $0\farcs201$ & $\ldots$ \\
\ F814W (WF3) & 375 & $0\farcs076$ & 1.08 & $0\farcs233$ & $0\farcs276$ & $2\farcs646$ \\
\ F555W (WF4) & 560 & $0\farcs072$ & 1.06 & $0\farcs253$ & $0\farcs307$ & $1100\arcsec$ \\
\ F814W (WF4) & 540 & $0\farcs109$ & 1.13 & $0\farcs257$ & $0\farcs295$ & $0\farcs996$ \\
\ Mean      & & & & $0\farcs222\pm0\farcs019$ & $0\farcs266\pm0\farcs019$ & $\ldots$ \\
N4449-27 \\
\ F555W (PC)  & 98  & $0\farcs167$ & 1.36 & $0\farcs213$ & $0\farcs223$ & $0\farcs248$ \\
\ F814W (PC)  & 80  & $0\farcs177$ & 1.55 & $0\farcs180$ & $0\farcs184$ & $0\farcs188$ \\
\ Mean      & & & & $0\farcs197\pm0\farcs017$ & $0\farcs204\pm0\farcs020$ & $0\farcs218\pm0\farcs030$ \\
N4449-47 \\
\ F555W (PC)  & 210 & $0\farcs138$ & 1.11 & $0\farcs316$ & $0\farcs376$ & $1918\arcsec$ \\
\ F814W (PC)  & 190 & $0\farcs139$ & 1.22 & $0\farcs237$ & $0\farcs261$ & $0\farcs468$ \\
\ Mean      & & & & $0\farcs277\pm0\farcs040$ & $0\farcs319\pm0\farcs058$ & $\ldots$ \\
N6946-1447 \\
\ F439W (PC)  & 285 & $0\farcs087$ & 0.94 & $0\farcs455$ & $0\farcs627$ & $\ldots$ \\
\ F555W (PC)  & 450 & $0\farcs080$ & 1.00 & $0\farcs336$ & $0\farcs432$ & $\ldots$ \\
\ F814W (PC)  & 820 & $0\farcs071$ & 1.05 & $0\farcs261$ & $0\farcs320$ & $145\arcsec$ \\
\ Mean      & & & & $0\farcs350\pm0\farcs056$ & $0\farcs460\pm0\farcs090$ & $\ldots$ \\
\enddata
\end{deluxetable}

\begin{deluxetable}{lrrr}
\tablecaption{\label{tab:kfits}Size measurements for the clusters based
  on King model fits within a $3\arcsec$ fitting radius. For N4214-13
  we have excluded the fit to the F336W image from the mean.}
\tablehead{  & FWHM & $r_t/r_c$ & \reff}
\startdata
N4214-10 \\
\ F336W (WF3) & $0\farcs174$ & 55 & $0\farcs339$ \\
\ F555W (WF3) & $0\farcs183$ & 36 & $0\farcs294$ \\
\ F814W (WF3) & $0\farcs203$ & 24 & $0\farcs269$ \\
\ Mean        &              &    & $0\farcs301\pm0\farcs020$ \\
N4214-13 \\
\ F606W (PC)  & $0\farcs082$ & 87 & $0\farcs197$ \\
\ F336W (WF3) & $0\farcs011$ & $10^9$ & $45\arcsec$ \\
\ F814W (WF3) & $0\farcs055$ & 713 & $0\farcs316$ \\
\ F555W (WF4) & $0\farcs056$ & 2566 & $0\farcs571$ \\
\ F814W (WF4) & $0\farcs087$ & 127  & $0\farcs251$ \\
\ Mean$^1$    &              &      & $0\farcs334\pm0\farcs083$ \\
N4449-27 \\
\ F555W (PC)  & $0\farcs142$ & 33 & $0\farcs193$ \\
\ F814W (PC)  & $0\farcs131$ & 41 & $0\farcs202$ \\
\ Mean        &              &    & $0\farcs198\pm0\farcs005$ \\
N4449-47 \\
\ F555W (PC)  & $0\farcs150$ & 52 & $0\farcs271$ \\
\ F814W (PC)  & $0\farcs122$ & 48 & $0\farcs214$ \\
\ Mean        &              &    & $0\farcs243\pm0\farcs029$ \\
N6946-1447 \\
\ F439W (PC)  & $0\farcs109$ & 225 & $0\farcs410$ \\
\ F555W (PC)  & $0\farcs089$ & 211 & $0\farcs324$ \\
\ F814W (PC)  & $0\farcs071$ & 178 & $0\farcs241$ \\
\ Mean        &              &    & $0\farcs325\pm0\farcs049$ \\
\enddata
\end{deluxetable}

\begin{deluxetable}{lrrrrrrrr}
\tablecaption{\label{tab:obslogkeck}Observation log for spectroscopic
  observations of the clusters. For HIRES observations, S/N estimates
  (per pixel) are given for order 11 ($\lambda\lambda6440-6520$\AA ).
  For NIRSPEC $H$ and $K$ settings the S/N are for orders 3 
  ($\lambda\lambda1.65-1.67\mu m)$ and 5 ($\lambda\lambda2.11-2.31 \mu m)$.}
\tabletypesize{\footnotesize}
\tablehead{Object & Instrument & Date & Slit & $\lambda/\Delta\lambda$ & Range & Orders & t(min) & S/N}
\startdata
N4214-10  & HIRES   & 2003-05-09 & $0\farcs861$ & 45000 & 5450--7800\AA & 20 & 150 & 38 \\
N4214-13  & HIRES   & 2003-05-08 & $0\farcs861$ & 45000 & 5450--7800\AA & 20 & 150 & 56 \\
N4449-27  & HIRES   & 2003-05-09 & $0\farcs861$ & 45000 & 5450--7800\AA & 20 & 150 & 15 \\
N4449-47  & HIRES   & 2003-05-08 & $0\farcs861$ & 45000 & 5450--7800\AA & 20 & 200 & 28 \\
N6946-1447 & NIRSPEC & 2002-07-13 & $0\farcs432$ & 25000 & H band       &  8 &  64 & 26 \\
N6946-1447 & NIRSPEC & 2002-07-13 & $0\farcs432$ & 25000 & K band       &  7 &  48 & 36 \\
\enddata
\end{deluxetable}

\begin{deluxetable}{lrr}
\tablecaption{\label{tab:tmpl_I_II}Supergiant template stars used for 
cross-correlation analysis. Spectral classifications are from the
Bright Star Catalogue and $M_V$ magnitudes were estimated using
Hipparcos parallaxes, neglecting interstellar absorption.}
\tablehead{Star  &  Spectral type & $M_V$}
\startdata
HR 2453 &  G5 Ib & $-1.7\pm0.9$ \\
HR 2615 &  K3 Ib & $\ldots$ \\
HR 3073 &  F1 Ia & $-2.1\pm0.7$ \\
HR 7456 &  G0 Ib & $-1.7\pm0.6$ \\
HR 7475 &  K4 Ib & $-2.9\pm1.2$ \\
HR 7892 &  K3 Ib & $-2.3\pm0.7$ \\
HR 2959 &  K3 II & $-1.6\pm0.3$ \\
HR 3229 &  G5 II & $-2.9\pm0.6$ \\
HR 7139 &  M4 II & $-2.9\pm0.3$ \\
HR 7479 &  G1 II & $-1.4\pm0.2$ \\
HR 7525 &  G5 II & $-3.0\pm0.2$ \\
HR 7823 &  F1 II & $-3.0\pm1.0$ \\
HR 7834 &  F5 II & $-2.8\pm0.3$ \\
HR 8003 &  K0 II & $-1.3\pm0.3$ \\
HR 8082 &  K0 II-III & $1.0\pm0.2$ \\
\enddata
\end{deluxetable}

\begin{deluxetable}{lrr}
\tablecaption{\label{tab:tmpl_III}Giant template stars used for 
cross-correlation analysis}
\tablehead{Star  &  Spectral type & $M_V$ }
\startdata
HR 5044 &  K  III & $0.41\pm0.16$ \\
HR 7413 &  K5 III & $-0.23\pm0.25$ \\
HR 7448 &  K4 III & $-1.54\pm0.43$ \\
HR 7509 &  M5 IIIa & $-0.33\pm0.30$ \\
HR 7566 &  M2 IIIa & $-1.27\pm0.23$ \\
HR 7583 &  K4 III & $-0.65\pm0.27$ \\
HR 7626 &  G9 III & $0.35\pm0.16$ \\
HR 7633 &  K5 II-III & $-2.27\pm0.31$ \\
HR 7811 &  G6 III & $-0.42\pm0.26$ \\
HR 7824 &  G8 III & $0.48\pm0.28$ \\
HR 7919 &  K2 III & $0.18\pm0.17$ \\
HR 7966 &  K3 III & $-1.59\pm0.56$ \\
HR 8011 &  K0 III & $-0.61\pm0.23$ \\
HR 8057 &  M1 III & $-0.09\pm0.35$ \\
HR 8066 &  K5 III & $-0.99\pm0.38$ \\
HR 8078 &  K0 III & $0.46\pm0.14$ \\
HR 8082 &  K0 II-III & $0.99\pm0.17$ \\
\enddata
\end{deluxetable}

 \begin{deluxetable}{lrrr}
 \tablecaption{\label{tab:vdmeas}Velocity dispersion measurements for
   each cluster as a function of template star luminosity class}
 \tablehead{ &       Ia/Ib    &       II       &      III}
 \startdata
 N4214-10    &  $4.1\pm1.4$ &  $4.8\pm1.4$ &  $5.4\pm1.0$ \\
 N4214-13    & $14.9\pm2.6$ & $14.8\pm2.8$ & $14.7\pm3.0$ \\
 N4449-27    &  $4.1\pm2.8$ &  $4.7\pm1.9$ &  $5.3\pm1.6$ \\
 N4449-47    &  $5.0\pm1.5$ &  $5.7\pm1.7$ &  $6.7\pm1.1$ \\
 N6946-1447  & \multicolumn{2}{c}{$8.8\pm1.5$} &   -          \\
 \enddata
 \end{deluxetable}

\begin{deluxetable}{lrrrrr}
\tablecaption{\label{tab:ppar}Adopted physical parameters for the
  clusters. $A_B$ values in this table are in addition to the
  foreground extinction from \citet{sch98}. For clusters in 
  NGC~4214 and NGC~4449 the $v_x$ values listed here are 
  averages of the values for LC II and LC III templates in 
  Table~\ref{tab:vdmeas}.}
\tabletypesize{\scriptsize}
\tablehead{  & N4214-10      & N4214-13       & N4449-27      & N4449-47 & N6946-1447}
\startdata
Distance (Mpc) & 2.8     &  2.8           &  3.9          &  3.9      & 6.0     \\
$v_x$ (km/s) & $5.1\pm1.0$ & $14.8\pm1.0$ & $5.0\pm1.0$ & $6.2\pm1.0$ &  $8.8\pm 1.0$ \\
$\reff$      & $0\farcs32\pm0\farcs01$ & $0\farcs22\pm0\farcs02$
	      & $0\farcs20\pm0\farcs02$ & $0\farcs28\pm0\farcs04$ & $0\farcs35\pm0\farcs06$ \\
$\reff$ (pc) & $4.33\pm0.14$ & $3.01\pm0.26$  & $3.72\pm0.32$ & $5.24\pm0.76$  & $10.2\pm1.6$ \\
$r_c$ (pc)   & $1.93\pm0.12$ & $0.64\pm0.06$  & $2.09\pm0.06$ & $1.45\pm0.01$  & $1.15\pm0.07$ \\
Log(age)     & $8.30\pm0.10$ & $8.30\pm0.10$  & $8.90\pm0.25$ & $8.45\pm0.10$  & $7.05\pm0.10$ \\
$A_B$ (mag)  & $0.09\pm0.04$ & $1.09\pm0.05$ & $-0.01\pm0.37$ & $0.48\pm0.05$  & $0.00$ \\
$M_V$        & $-10.22$      & $-11.68$       & $-9.61$       & $-10.74$   & $-13.19$ \\
$\mvir$ ($\times10^5 \msun$) 
	     & $2.6\pm1.0$ & $14.8\pm2.4$  & $2.1\pm0.9$ & $4.6\pm1.6$ & $17.6\pm5$ \\
$M/L_V$      & $0.25\pm0.10$ & $0.38\pm0.06$ & $0.36\pm0.15$ & $0.28\pm0.10$ & $0.11\pm0.03$ \\
$\mu_0$ ($V$ mag arcsec$^{-2}$) & $14.9\pm0.2$ & $11.8\pm0.2$ & $15.5\pm0.1$ & $14.2\pm0.1$ & $12.2\pm0.2$ \\
$\rho_{\rm hmr}$ (M$_{\odot}$ pc$^{-3}$) & $340\pm140$     & $6000\pm1300$      & $420\pm190$  & $340\pm150$  & $180\pm80$     \\
$\rho_0$ (M$_{\odot}$ pc$^{-3}$) & $(2.5\pm1.0)\times10^3$ & $(1.9\pm0.6)\times10^5$  & $(1.9\pm0.8)\times10^3$ & $(6.8\pm2.4)\times10^3$ & $(2.3\pm0.8)\times10^4$ \\
\enddata
\end{deluxetable}

\end{document}